\documentclass[%
rsi,
amsmath,
amssymb,
reprint,%
]{revtex4-1}

\usepackage{graphicx}
\usepackage{amsfonts}
\usepackage{booktabs}
\usepackage{mathtools}
\usepackage{pgfplots}
\usepackage{geometry}
\usepackage{enumerate}
\usepackage{esvect}
\usepackage{footnote}
\usepackage{listings}
\usepackage{xcolor}
\usepackage{enumitem}
\usepackage{tocloft}
\usepackage{textcomp}
\usepackage{braket}
\usepackage{nicefrac}
\usepackage{nccmath}
\usepackage{bm}
\usepackage{import}
\usepackage{dpfloat}
\usepackage{nicefrac}
\usepackage{subcaption}
\usepackage{float}
\usepackage{afterpage}
\usepackage{multirow}
\usepackage{siunitx}
\usepackage{tikz}
\usepackage{wasysym}
\usepackage[ruled]{algorithm2e}
\DeclarePairedDelimiter\abs{\lvert}{\rvert}%
\usepackage{graphicx}
\newcommand{\CC}{%
    {\settoheight{\dimen0}{C}C\kern-.05em \resizebox{!}{\dimen0}{\raisebox{\depth}{++}}}}
\newcommand{\CCC}{%
    {\settoheight{\dimen0}{C}C/C\kern-.05em \resizebox{!}{\dimen0}{\raisebox{\depth}{++}}}}
\newcommand{\CS}{%
    {\settoheight{\dimen0}{C}C\kern-.05em \resizebox{!}{\dimen0}{\raisebox{\depth}{\#}}}}

\draft

\begin{document}
    \suppressfloats 
    
    \title[Automated LIDT testing applied to a ferroelectric SLM]{Automated laser induced damage threshold testing applied to a ferroelectric spatial light modulator}
    
    \author{Peter J. Christopher}
    \email[]{pjc209@cam.ac.uk}
    \homepage[]{www.peterjchristopher.me.uk}
    \affiliation{Centre of Advanced Photonics and Electronics, University of Cambridge, 9 JJ Thomson Ave, Cambridge, UK, CB3 0FF}%
    \author{Nadeem Gabbani}
    \affiliation{Institute for Manufacturing, University of Cambridge, 17 Charles Babbage Rd, Cambridge, UK, CB3 0FS}%
    \author{William O'Neill}
    \affiliation{Institute for Manufacturing, University of Cambridge, 17 Charles Babbage Rd, Cambridge, UK, CB3 0FS}%
    \author{Timothy D. Wilkinson}
    \affiliation{Centre of Advanced Photonics and Electronics, University of Cambridge, 9 JJ Thomson Ave, Cambridge, UK, CB3 0FF}%
    
    \date{\today}%

    \begin{abstract}
    	Laser Induced Damage Thresholds~(LIDTs) are a measure of the level of fluence an optical component may be expected to handle without observable damage. In this work we present an automated system for measurement of LIDTs for a wide range of components and include details of the data handling, image processing and code required.
        
        We then apply this system to LIDT measurements of a commercial HDP-1280-2 'BlueJay' ferroelectric display finding a LIDT of $9.2~\si{\watt\per\centi\metre\squared} \diameter 27 \si{\micro\metre}$, $5.5~\si{\watt\per\centi\metre\squared} \diameter 150 \si{\micro\metre}$ and $3.2~\si{\watt\per\centi\metre\squared} \diameter 3.1 \si{\milli\metre}$ with wavelength $1090\pm5\si{\nano\meter}$. Finally, the quality of the results obtained are discussed and conclusions drawn.            
    \end{abstract}
    
    \pacs{}%
    
    \maketitle
           
    \section{Introduction}
        
    Laser Induced Damage Thresholds~(LIDTs) are an internationally standardised way of quantifying the threshold laser fluence required to cause damage in optical elements. In this paper we present an automated system for LIDT testing in accordance with the ISO 11254 and ISO 21254 standards~\cite{ISO11254, ISO21254}. 
    
    As part of ongoing research into the power handling capabilities of Spatial Light Modulators (SLMs) we present an automated system for measurement of damage threshold values. We demonstrate this for a commercial Liquid Crystal on Silicon~(LCoS) device under Continuous Wave~(CW) laser illumination. This substrate was chosen as a significant quantity of devices were available along with detailed accompanying power handling measurements for comparison. 
    
    For CW power sources, the optic is exposed at $10$ locations to a laser of known beam diameter and power. The result is then examined under a high magnification optical microscope for visible damage. The laser power is varied between measurements with the LIDT being taken as the highest laser power for which damage is not observed on any of the $10$ exposure sites. 'Damage' is here defined according to the ISO definition as any detectable change in the substrate. 
    
    As bulk heating is assumed to be the primary mechanism for damage under CW exposure,~\cite{WOOD1998517} LIDTs are often quoted with the associated beam diameter.~\cite{2014Ldio,Bloembergen:73} We here give the LIDT in terms of power per area or $\si{\watt\per\centi\metre\squared}$ for a $\nicefrac{1}{e^2}$ beam diameter $\diameter$ given in $\si{\micro\metre}$ or $\si{\milli\metre}$. LIDTs may also be given in terms of the effective area equal to the ratio of laser power to maximum power density.~\cite{ISO11254,PhysRevLett.91.127402}
    
    The ISO standards do not require a specific beam profile and only maximum beam intensity and beam diameter are required. In the case discussed here detailed manufacturer specifications were available for the beam diameter. For Gaussian illumination, the peak power is approximately $2 \times$ the equivalent power of an equivalent uniformly distributed beam.~\cite{WOOD1998517}
    
    The primary motivation for this work is the automation of a task for improvement in speed and reduction in human error. We begin by presenting the experimental setup and automation arrangements with a focus of methodology. The system is then validated using an LCoS device as a test case. Finally, the measured response is discussed and conclusions are drawn. 
    
    \section{Experimental Setup}
    
    Manual LIDT measurement is straightforward requiring only a known light source, a means of attenuation, a substrate and a microscope. A 'plug-and-play' automated system requires little further work. Some devices such as polarising filters require light of known polarisation and the damage thresholds for components can range significantly. 
    
    \begin{figure}
        \centering
        {\includegraphics[width=1.0\linewidth]{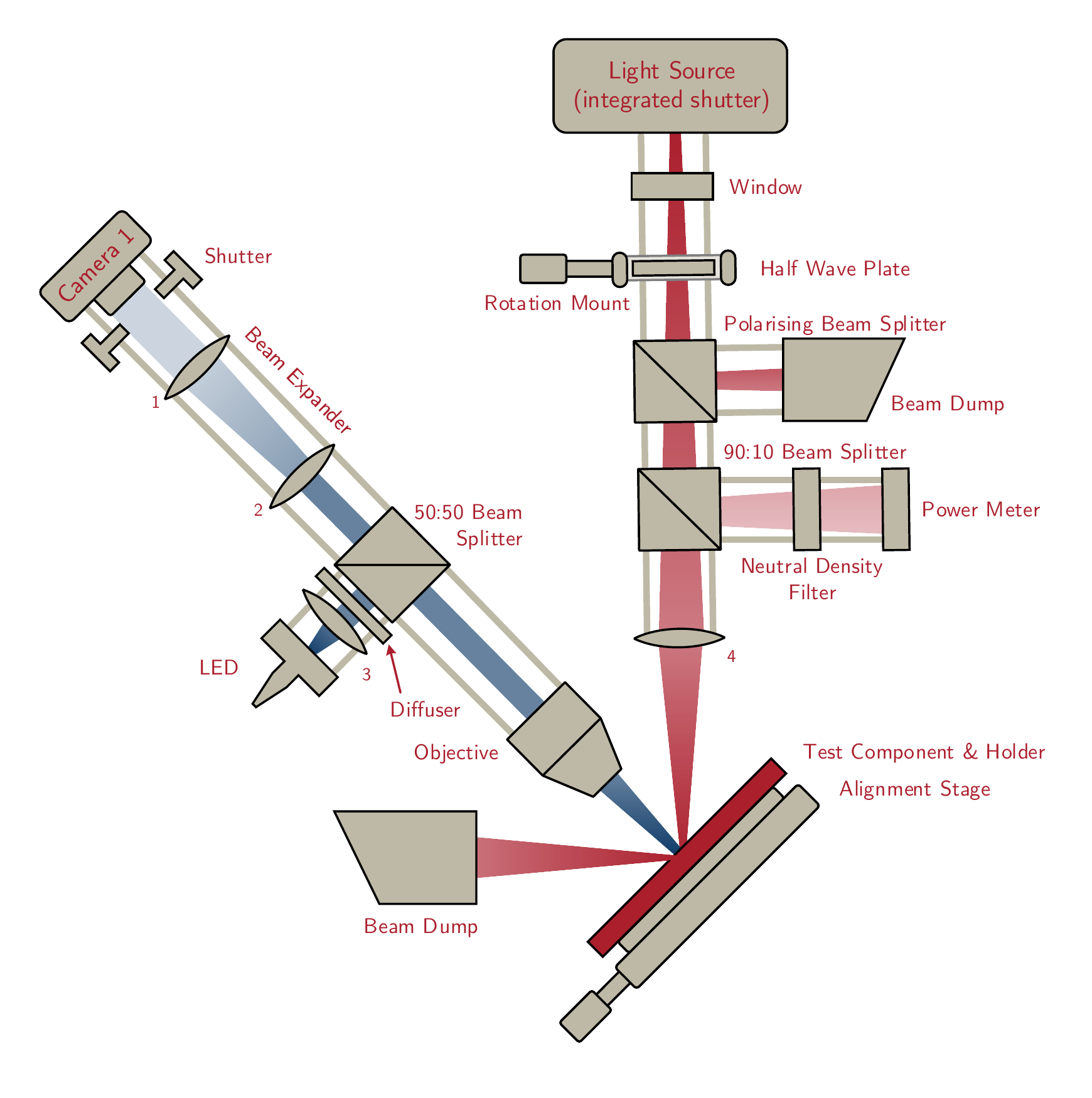}}
        \caption{Experimental schematic}
        \label{fig:schematic}
    \end{figure}
    
\begin{figure}
        \centering
        {\includegraphics[width=1.0\linewidth]{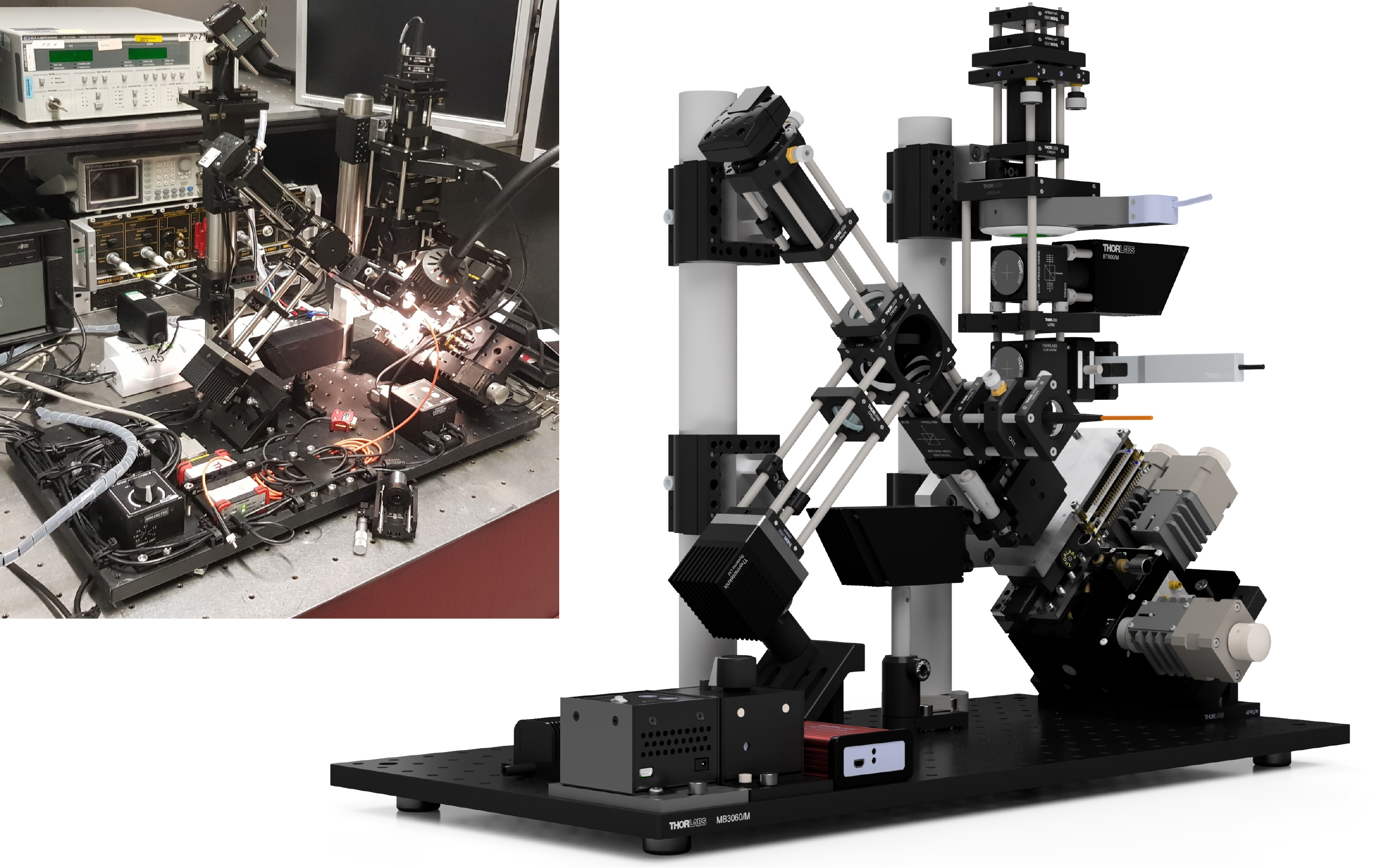}}
        \caption{System design (main) and implementation (inset)}
        \label{fig:combo}
    \end{figure}

    Figure~\ref{fig:schematic} shows the schematic of the system designed for automating this process. The Computer Aided Design~(CAD) design is shown in Figure~\ref{fig:combo} along with its real-world implementation. A 'plug-and-play' approach is taken for the laser source which can include a range of directly cage mountable sources including diodes as well as fibre launched light sources.
    
    The laser beam - red in Figure~\ref{fig:schematic} - is passed vertically downwards through a window and adjustable linear polariser. Adjustment of the polariser relative to the fast axis of the polarising beam splitter (PBS) allows for intensity control in elliptical beams and ensures a know polarisation on the sample. A 90:10 or higher beam splitter extracts a portion of the power for power measurement and a switchable neutral density (ND) filter ensures compatibility with a wide range of intensities. Len 4 acts as a telescope with the distance between the lens and the stage defining the incident beam spot. Integration with Zemax allows the control system to automate this process. Any reflected light is captured by the beam dump.
    
    The microscope system - blue in Figure~\ref{fig:schematic} - operates by passing a white light LED source through an objective and imaging the back reflected light. This allows for real time measurement of substrate degradation.
    
    In order to ensure maximum flexibility, all components in the system are designed to be modular and interchangeable. 

    \section{Automation, Control and Operation}
    
    A control suite for the system was developed in \CS{} and \CCC{} based on the HoloGen framework~\cite{hologen}. This is capable of automating the entire alignment, characterisation and metrology process with a minimum of initial user input. 
    
    \subsection{Source Calibration}
    
    There are three automated calibration procedures for the source measuring power, stability and ellipticity.
    
    \subsubsection{Source Power Calibration}
    
    Calibration of the laser source power is straight forward provided the power sensor used is of known properties. The waveplate and polarising beam splitter are aligned with parallel fast axes and the response curve of source driving voltage to measured power is taken. Aligning the fast axis of the laser at $45^{\circ}$ to the polarising beam splitter and repeating the measurement allows a second response curve to be measured. The combined response of the laser is equal to the sum of these measurements and allows us to determine laser power without removing the polarising beam splitter or half waveplate.
    
    \subsubsection{Source Stability Calibration}
    
    The stability of the laser source can be determined simply by holding the source at constant driving voltage and recording the change in measured power over a period of time, in this case taken as a period of 8 hours. Taking sufficient measurements allows a least squares fit to a gaussian distribution in order to calculate the FWHM stability. In the application discussed below, stability was sufficient to ignore it from LIDT calculations.
    
    As before, systems incorporating the waveplate and polarising beam splitter require two measurements at $45^{\circ}$ in order to fully understand stability behaviour.
    
    \begin{figure}
        {\includegraphics[width=\linewidth,page=1]{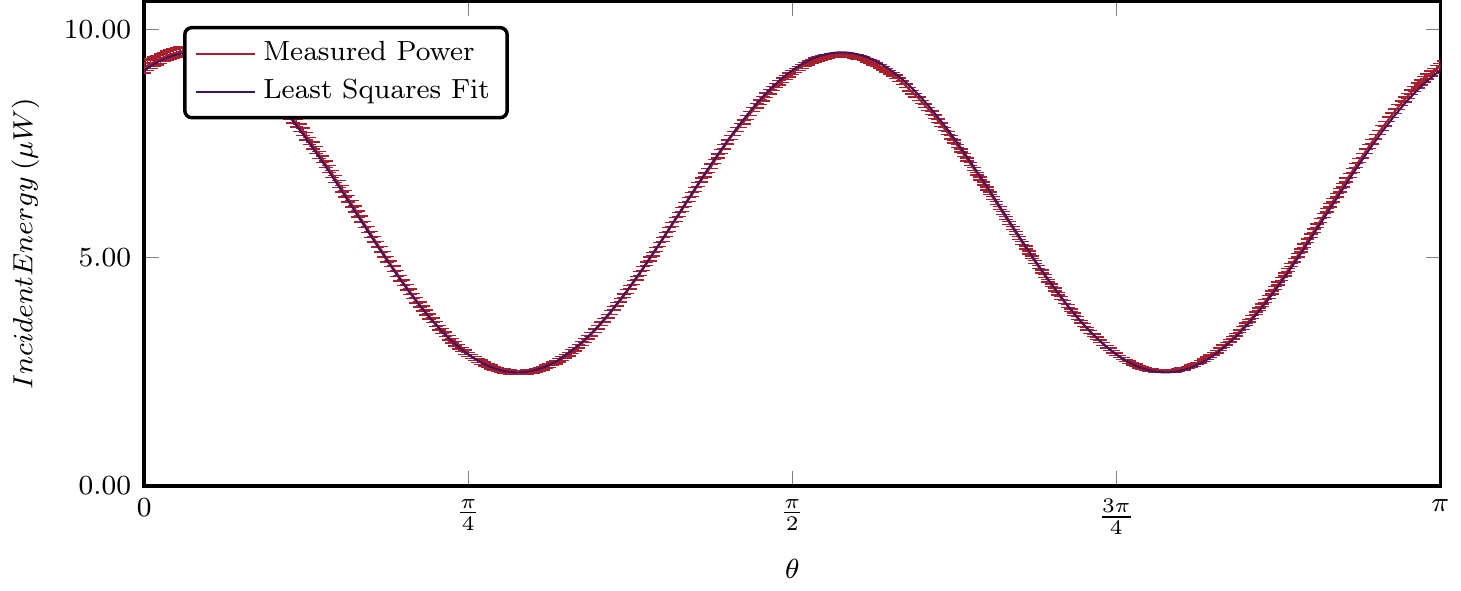}}
        \caption[Wave Plate Effect]{Power incident on the power meter vs waveplate angle for the system shown in Figure~\ref{fig:schematic} and a 10mW solid state laser}
        \label{fig:waveplate}
    \end{figure}
    
    \subsubsection{Source Ellipticity Calibration}
    
    Slightly more involved is the source ellipticity calibration. For an arbitrary elliptical polarisation $E=[A, B e^{i\delta}]$ passed through a polariser of variable orientation, the minimum and maximum values are given by
    
    \begin{align} \label{jones3}
    &E_{\psi}                    = \\
    &E\sqrt{A^2\cos^2\psi + B^2\sin^2\psi + AB\cos\delta\sin{2\psi}} \nonumber\\
    &E_{\psi \pm \frac{\pi}{2}}  = \\
    &E\sqrt{A^2\sin^2\psi + B^2\cos^2\psi - AB\cos\delta\sin{2\psi}} \nonumber
    \end{align}
    
    where $A$, $B$, $E$ and $\delta$ are scalar constants, $\sqrt{A^2+B^2}=1$ and where $\psi$ is given by
        
    \begin{equation} \label{jones2}
        \psi=\frac{1}{2}\tan^{-1}{\left(\frac{2AB\cos{\delta}}{A^2-B^2}\right)}
    \end{equation}
    
    This gives a relationship for measured intensity $I$ of 
        
    \begin{widetext} 
        \begin{equation} \label{jones7}
        I=
        \underbrace{C\frac{\left(A^2+B^2\right)}{2}}_\text{Constant Term} +
        \underbrace{C\sqrt{\frac{\left(A^2-B^2\right)^2+A^2B^2\cos{\delta}^2}{2}}}_\text{Amplitude Term}\times  
        \underbrace{\sin\left(4\theta_0+4\theta+\tan^{-1}\left(\frac{\left(A^2-B^2\right)}{2AB\cos{\delta}}  \right)  \right)}_\text{Frequency Term} 
        \end{equation}
    \end{widetext}
    
    where constant $C$ incorporates the scaling and loss terms of the system and $\theta$ is the angle subtended by the waveplate. 
    
    When the waveplate is initially mounted at a non-zero angle $\theta_{0}$ and the source is mounted with unknown orientation the waveplate is rotated through 360 degrees and the incident powers recorded. An example is shown in Figure~\ref{fig:waveplate}. Linear regression then allows for determination of source properties from which the ellipticity can be determined.
    
    \subsection{Computer Vision}
    
    The initial focus of the microscope sub-system is set by the user. A basic software autofocus implementation is used with an integrated Zemax model of the objective lens system used to inform z-axis adjustments on the alignment stage. The power of the illumination LED is controlled to ensure good image white-balance and contrast and reduce post processing.
    
    To automate the damage observation process, a control image is taken before the start of each test. After each test the recorded image $I_{i}$ for measurement $i$ is compared to the control image $I_{0}$ using a normalised mean squared error $E_{MSE}$ where
    
    \begin{widetext} 
        \begin{equation}
    E_{MSE}(I_{0},I_{i}) = 
    \frac{1}{N_x N_y}\sum_{x=0}^{N_x-1}\sum_{y=0}^{N_y-1} \left[k\abs{I_{0}(x,y)} -  \abs{I_{i}(x,y)}\right]^2 
    \quad\textit{where} \quad k=\sum_{x=0}^{N_x-1}\sum_{y=0}^{N_y-1}\frac{{\abs{I_{i}(x,y)}}^2}{{\abs{I_{0}(x,y)}}^2}
        \end{equation}
\end{widetext}
    
    and $N_x$ and $N_y$ are the respective $x$ and $y$ resolutions. A suitable cutoff value for $E_{MSE}$ can then be taken. All captured images are preserved to allow for manual confirmation if required.
    
    \begin{figure}
        {\includegraphics[width=\linewidth,page=1]{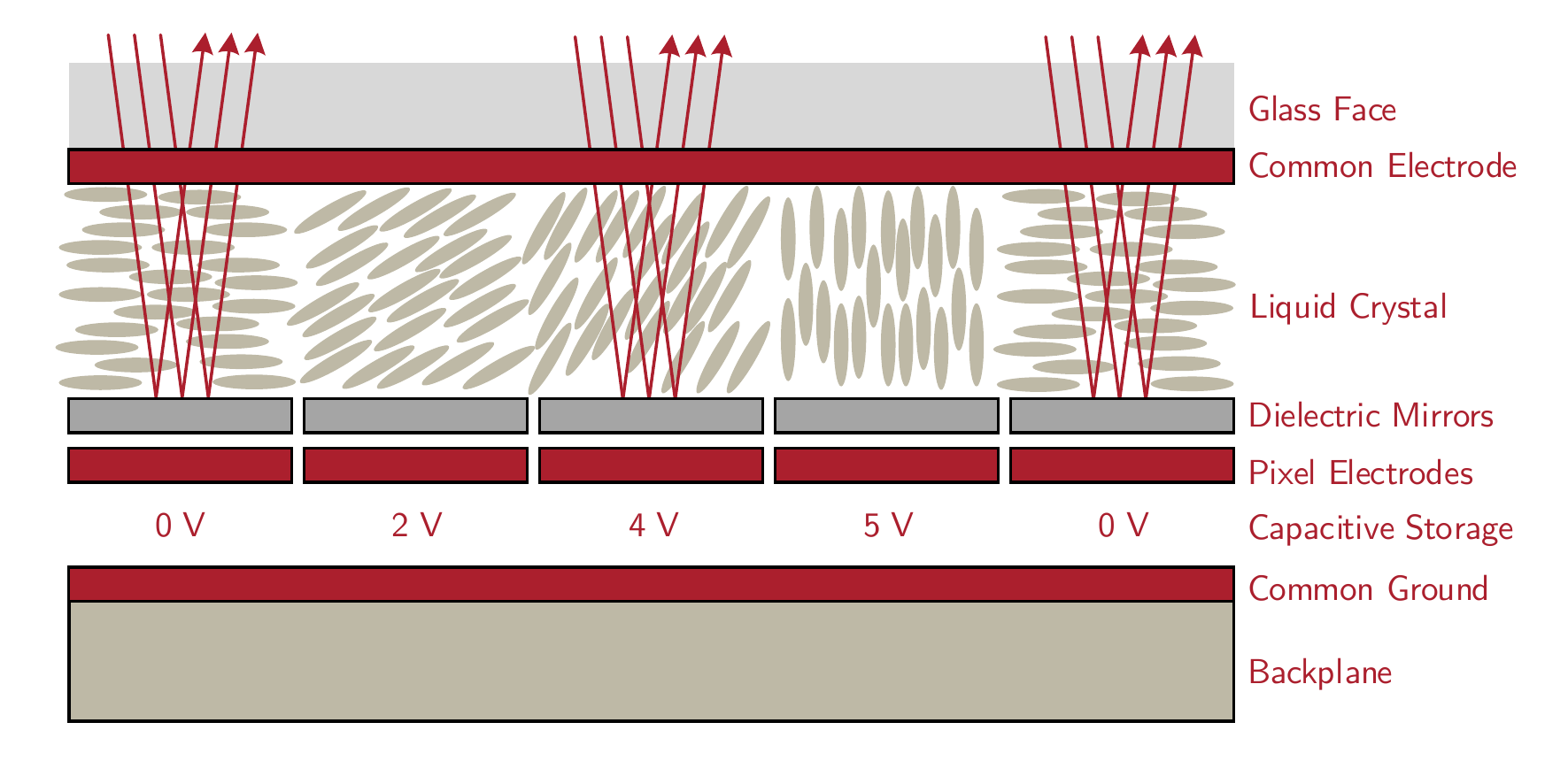}}
        \caption{Structure of a spatial light modulator}
        \label{fig:slm}
    \end{figure}
    
    \begin{figure*}[htbp]
        {\includegraphics[width=\linewidth,page=1]{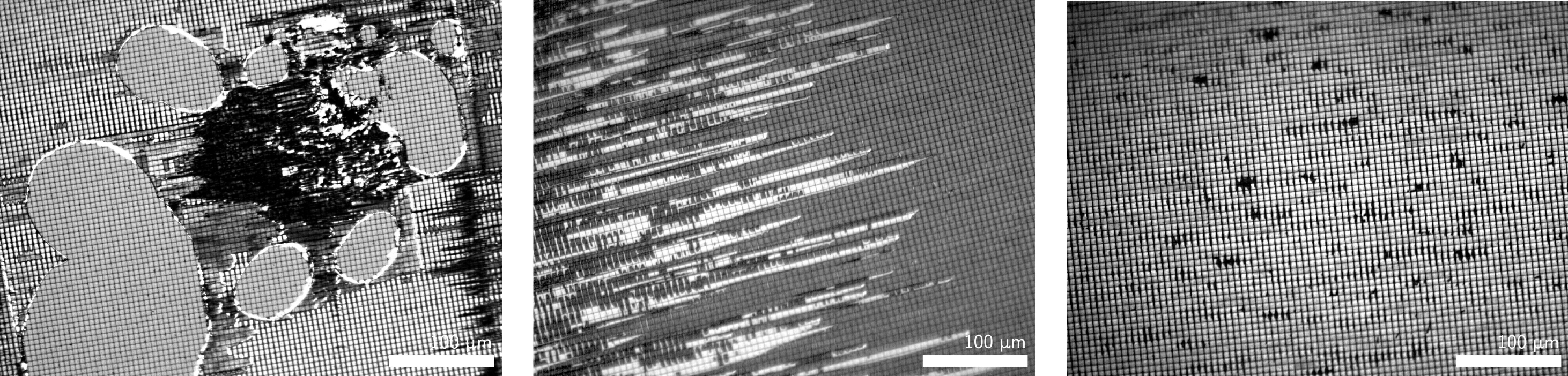}}
        \caption{Optical microscope images of substrate damage using an Evolution MP 5.0 camera and Olympus BX60 microscope}
        \label{fig:capture}
    \end{figure*}

    The system parameters are set by the user in a JSON format. These define the volume of operation for the stage as well as testing area on the component and initial power values for testing.
    
    \section{Validation}
    
    In order to validate the system we used NENIR30A ND filters from ThorLabs as they are low cost and have well defined damage thresholds under CW illumination. ThorLabs specify the NENIR30A as having a LIDT of $25~\si{\watt\per\centi\metre\squared} \diameter 62 \si{\micro\metre}$ at $1064$nm while we measured a LIDT of $29.6~\si{\watt\per\centi\metre\squared} \diameter 70 \si{\micro\metre}$ at $1090$nm. 
    
    \section{Demonstration} \label{results}
    
    The experimental rig discussed so far was designed as part of ongoing research into SLM power handling capabilities and we demonstrate our system using a number of HDP-1280-2 'BlueJay' ferroelectric displays. These have a resolution of $1280\times 1280$ pixels and a package size of 11\si{\milli\metre} by 25\si{\milli\metre}.
    
    As SLMs are multi-level devices, Figure~\ref{fig:slm}, we take the definition of 'damage' to include any visible change in the device rather than simply visible change in the substrate.
    
    As interest was in Near Infrared~(NIR) behaviour, a $200W$ $1090\pm5\si{\nano\meter}$ fibre laser source from  was used. This is delivered to the system through a multi-mode fibre.
    

    \section{Results and Discussion}
    
    The automated system ran $\approx 350$ tests for a number of spot sizes. The operator time for testing was under $35$ minutes with a combined automated runtime of $6$ hours. It is estimated that an entire operator day would be required in an equivalent manual system.
    
    As can be expected, the measured maximum power was higher for smaller gaussian spot sizes with LIDTs of $9.2~\si{\watt\per\centi\metre\squared} \diameter 27 \si{\micro\metre}$, $5.5~\si{\watt\per\centi\metre\squared} \diameter 150 \si{\micro\metre}$ and $3.2~\si{\watt\per\centi\metre\squared} \diameter 3.1 \si{\milli\metre}$ being measured at $1090\pm5\si{\nano\meter}$. This is presumed to be due to bulk heating. 
    
    A number of failure paradigms were observed with some extremal cases shown in Figure~\ref{fig:capture}. Figure~\ref{fig:capture}~(left) shows liquid crystal breakdown under prolonged exposure. Figure~\ref{fig:capture}~(centre) shows delamination of the liquid crystal from the glass without substrate damage and Figure~\ref{fig:capture}~(right) shows direct substrate damage. The captured microscope images were manually inspected with only one image being classified differently by the human operator and computer vision system.
    
    While not unexpected, there was no observed difference in the LIDT against polarisation parallel or perpendicular to the SLM major axis.
    
    \section{Conclusion}
    
    This work has presented a fully automated system for Laser Induced Damage Threshold testing of substrates using only commercial off-the-shelf components. The setup requires $<10\%$ of the operator time required for the equivalent manual system and reduces the manual error sources. 
    
    The system was demonstrated by testing a Liquid Crystal on Silicon (LCoS) device. LIDTs of $9.2~\si{\watt\per\centi\metre\squared} \diameter 27 \si{\micro\metre}$, $5.5~\si{\watt\per\centi\metre\squared} \diameter 150 \si{\micro\metre}$ and $3.2~\si{\watt\per\centi\metre\squared} \diameter 3.1 \si{\milli\metre}$ were found for the active device face with an excitation wavelength of $1090\pm5\si{\nano\meter}$.

    \begin{acknowledgments}
    
    The authors would like to thank the Centre for Advanced Photonics and Electronics~(CAPE) technical team for their tireless efforts in helping construct this system. In particular we would like to thank Mr Stephen Drewitt, Mr Ady Ginn, Mr Mark Barnett, Mr Joe Smith and Mr Dave Edwards. The authors would also like to thank the Centre for Doctoral Training in Ultra Precision Engineering (CDT-UP) and the Engineering and Physical Sciences Research Council (EPSRC) for financial support (EP/L016567/1). 
    \end{acknowledgments}
     
    \bibliography{references}

\end{document}